\documentclass{aa}
\usepackage{graphicx}
\usepackage{tabularx}
\usepackage{rotating}
\usepackage{txfonts}
\usepackage{comment}
\usepackage{lscape}
\usepackage{threeparttable}
\usepackage{xcolor}
\usepackage{subfigure}
\usepackage{caption}
\usepackage{float}
\usepackage[switch]{lineno}
\usepackage{ulem}

\newcommand{\apx}{$\sim$}

\newcommand{\target}{B2~0258+35}
\newcommand{\eg}[1]{\citep[e.g.][]{#1}}
\newcommand{\kmps}{km~s$^{-1}$}
\newcommand{\p}[1]{$^{-#1}$}
\newcommand{\pp}[1]{$^{#1}$}

\newcommand{\halpha}{H$\alpha$}

\newcommand{\beq}{\begin{equation}}
\newcommand{\eeq}{\end{equation}}

\newcommand{\Msun}{M$_\odot$}

\begin{document}

   \title{Cold gas bubble inflated by a low-luminosity radio jet}

   \author{Suma Murthy\inst{1},
          Raffaella Morganti\inst{2,3},
          Tom Oosterloo\inst{2,3},
          Dipanjan Mukherjee\inst{4},
          Suude Bayram\inst{5,2},
          Pierre Guillard\inst{6,7},
          Alexander Y. Wagner\inst{8},
          Geoffrey Bicknell\inst{9}
         }

   \institute{Joint Institute for VLBI ERIC, Oude Hoogeveensedijk 4, 7991 PD Dwingeloo, The Netherlands. \\
                \email{murthy@jive.eu}
        \and
             ASTRON, The Netherlands Institute for Radio Astronomy, Oude Hoogeveensedijk 4, 7991 PD Dwingeloo, The Netherlands.
        \and
            Kapteyn Astronomical Institute, University of Groningen, P.O. Box 800, 9700 AV Groningen, The Netherlands.
        \and
            Inter-University Centre for Astronomy and Astrophysics, Post Bag 4, Pune - 411007, India.
        \and
        Astronomy and Space Sciences Department Faculty of Science, Erciyes University 38039, Kayseri, T\"urkiye.
        \and
            Sorbonne Universit\'e, CNRS, UMR 7095, Institut d’Astrophysique de Paris, 98 bis bd Arago, 75014 Paris, France.
        \and
            Institut Universitaire de France, Minist\`ere de l’Enseignement Sup\'erieur et de la Recherche, 1 rue Descartes, 75231 Paris Cedex F-05, France.
        \and
            Center for Computational Sciences, University of Tsukuba, 1-1-1 Tennodai, Tsukuba, Ibaraki, 3058577
        \and
            Australian National University, Research School of Astronomy and Astrophysics, Cotter Rd., Weston, ACT 2611, Australia
}

   \date{Received: November 25, 2024, accepted January 13, 2025}


 \abstract{We present NOEMA CO(2-1) observations of a nearby, young, low-luminosity radio source, \target. Our earlier CO(1-0) study had shown the presence of strong jet-ISM interaction and a massive molecular gas outflow involving 75\% of the circumnuclear gas. Our follow-up CO(2-1) observations have revealed even more complex gas kinematics, where the southern radio jet is driving out molecular gas in the form of a swiftly expanding bubble, with velocities up to almost 400 \kmps. We found highly elevated CO(2-1)/CO(1-0) line ratios for the gas belonging to the bubble and also further away from the radio jets. Previous observations have shown that the active galactic nucleus (AGN) in the host galaxy, NGC~1167, is in a very low-accretion state. Thus, we attribute the high line ratios to the high gas excitation caused by the jet--ISM interaction. The radio jets, despite exhibiting a relatively low luminosity ($1.3 \times 10^{44}$ erg s\p{1}), are solely responsible for the observed extreme gas kinematics. This is one of the clearest detections of an expanding cold gas bubble in such a type of source, showing that the jets are affecting both the kinematics and physicals conditions of the gas. Our study adds to the growing store of evidence that low-luminosity radio sources can also affect the kinematics and physical conditions of the cold gas, which fuels star formation, in their host galaxies to a significant extent. Hence, such sources should be considered in models seeking to quantify feedback from radio AGN.}


\titlerunning{Cold gas bubble in B2~0258+35}
\authorrunning{Murthy et al.}

\maketitle
\section{Introduction} \label{sec:intro}

The radio jets arising from active supermassive black holes are becoming increasingly relevant in the context of galaxy evolution given their role in impacting the gas within their host galaxies. A growing number of theoretical and observational studies have shown that these jets, starting from a very early stage of their evolution, interact with the interstellar medium (ISM) and make the gas highly turbulent, driving multi-phase outflows and   enhancing the line ratios of different molecular transitions \citep[e.g.][]{Holden24, Morganti23a, Morganti21, Oosterloo19, Morganti15, Garcia-Burillo07}.

\begin{figure*}
\centering
    \includegraphics[width=6cm]{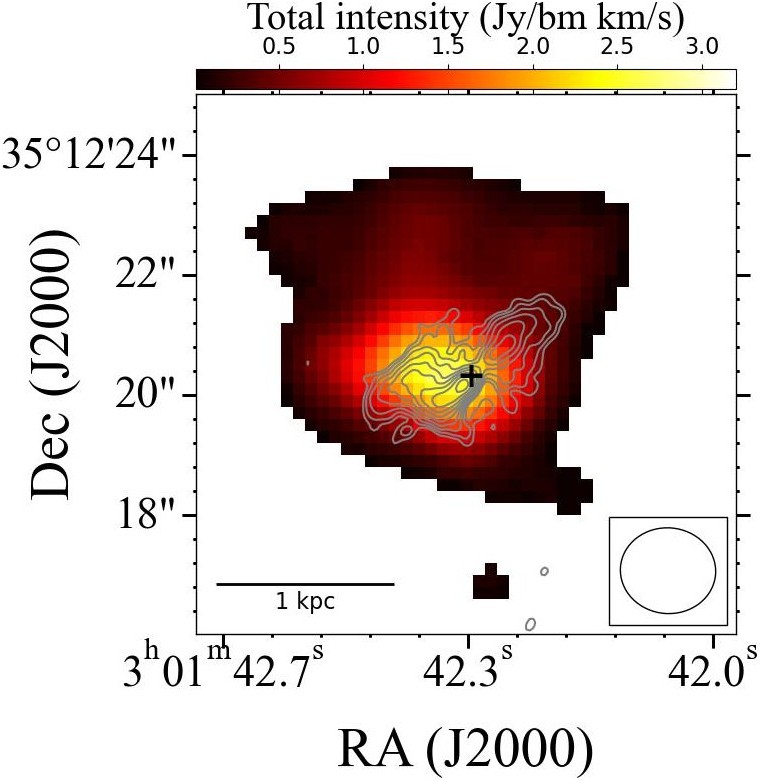}
    \includegraphics[width=6cm]{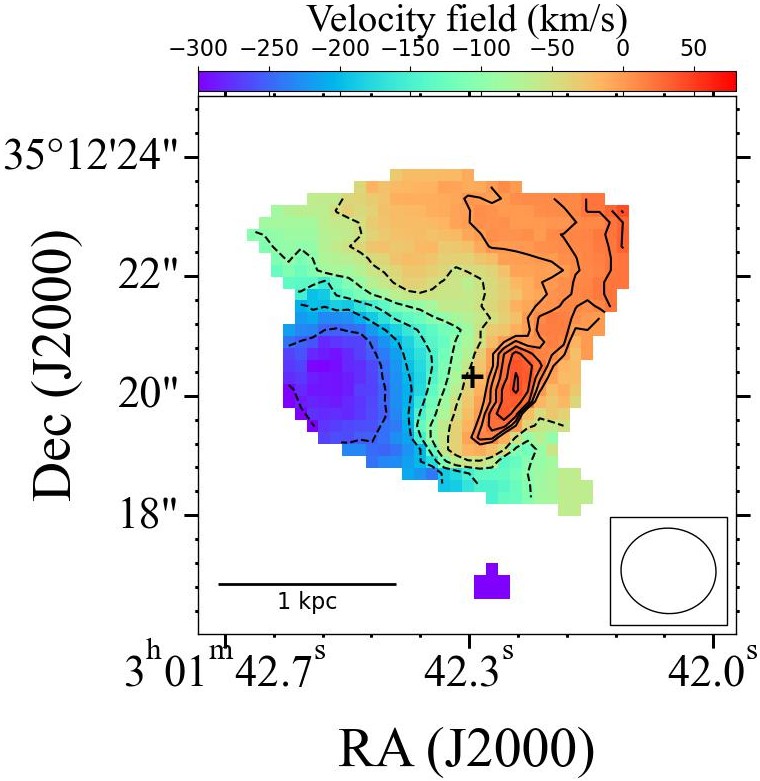}
    \includegraphics[width=6cm]{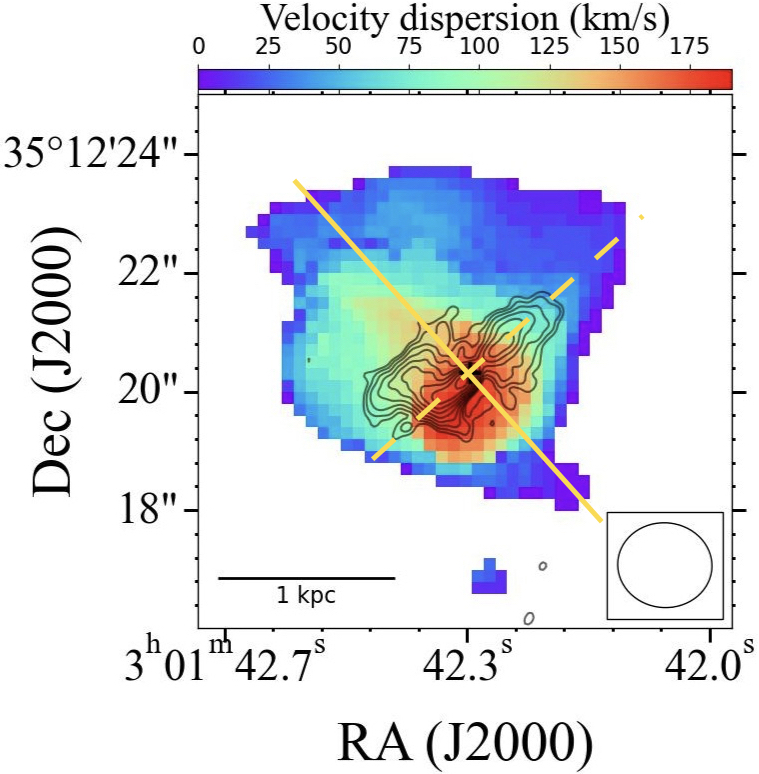}

    \caption{Moment maps of the CO(2-1) emission. \textbf{Left:} Total intensity (moment-0) map, \textbf{Centre:} Velocity field (moment-1 map). The dashed contours represent the blue-shifted velocities and the solid contours represent the redshifted velocities, and \textbf{Right:} Velocity dispersion (moment-2) map. The yellow solid line indicates the radio axis and the yellow dashed line indicates the axis is perpendicular to the radio axis. These two correspond to the axes of the position-velocity slices shown in Fig. \ref{fig:pvd}a,c and Fig. \ref{fig:pvd}b, respectively. The 8.4 GHz radio continuum from \citet{Giroletti05} is shown in grey and black contours in moment-0 and moment-2 maps, respectively. The radio core is marked with a black cross in all three images.
    }
    \label{fig:moms}
\end{figure*}

Low-luminosity radio active galactic nuclei (AGN; of $<10^{23}$ W Hz\p{1}) are of particular interest for such studies since 30\% of massive galaxies host such radio sources. Indeed, there is a rising number of studies in this direction, demonstrating that low-luminosity radio sources affect the ISM in the ways mentioned above \citep[e.g.][]{Morganti15, Dasyra16, Girdhar22, Girdhar24,Venturi23,Audibert23, Ulivi24}.

However, often in such studies, the effect of the optical AGN on the ISM cannot be disentangled from that of the radio jets and the sources where the observed impact on the ISM can solely be attributed to low-power radio jets are rare. In cases where the optical AGN is indeed too weak to have any significant impact on the gas, the radio jet-ISM interaction is often unresolved, thereby limiting the information that can be derived regarding the morphology and kinematics of the affected gas \citep[e.g.][]{Alatalo11, Combes13, Fabbiano18b}. Ideally, to understand the impact, we should study sources where the jet-ISM interaction can be spatially resolved and is due to radio jets alone. At the moment, there are very few such sources that can be studied in detail \citep[e.g.][]{Ruffa22}, with \target\ being one of them.

\target\ is a low-luminosity (L$_{\rm1.4~GHz} = 2.1 \times 10^{23}$ W Hz\p{1}), restarted radio AGN. It has a compact steep spectrum (CSS) radio source that is \apx 1 kpc in size in the centre \citep{Giroletti05} and exhibits extremely low-surface-brightness radio lobes on larger scales \citep[\apx 240 kpc;][]{Shulevski12, Brienza18}. The central radio source is young with an estimated age of 0.4 -- 0.9 Myr \citep{Brienza18}. It consists of a faint core and two jets. The northern jet is relatively fainter and has a diffuse morphology. The brighter southern jet consists of a radio hot spot \apx 0.5 kpc from the core and bends sharply at the location of this hot spot \citep[Fig. \ref{fig:moms} black contours;][]{Giroletti05}. 

The host galaxy, NGC~1167, is rich in atomic \eg{Struve10c, Murthy19} and molecular gas \eg{ Prandoni07, O'Sullivan15, Murthy22b}. In our earlier NOEMA  study, where we could spatially resolve the CO(1-0), we found that the kinematics of the cold molecular gas in the  few central kpc are different from the large-scale rotation of the galaxy \citep{Murthy22b}. In the entire circumnuclear region, the gas is highly turbulent due to the jet-ISM interaction. The interaction is strongest along the southern radio jet, where the jet drives a massive (\apx 10\pp{6} \Msun) outflow. This outflow constitutes 75\% of the cold molecular circumnuclear gas and is localised 500 pc away from the core along the southern radio jet, where the jet bends sharply. Also, the region with high velocity dispersion ($>$ 200 \kmps) extends in a  perpendicular direction to the radio axis \citep[see Figs. 1 and 2 in][]{Murthy22b}.  Furthermore, the \textit{Chandra} X-ray follow-up study of NGC\,1167/\target\ \citep{Fabbiano22} detected soft X-ray emission enveloping the southern radio jet as well as extending perpendicular to the radio axis, coinciding with the location of the CO outflow, and indicating the presence of dense X-ray emitting gas and strong shocks arising due to jet--ISM interaction. However, these conclusions were limited by small-number statistics due to low photon counts. With the present study, we are able to place this scenario of jet--ISM interaction on a firmer footing. 

The faint hard X-rays detected by \textit{Chandra} and the stringent NuSTAR upper limits on the harder X-ray emission suggest that the AGN in NGC~1167 is in a very low-accretion state. This confirms the earlier suggestions based on the optical properties. Optically, NGC\,1167/\target\ is a LINER galaxy \citep{Ho09}, with a bolometric luminosity for the AGN in the range of $(0.3 - 7) \times 10^{42}$ erg s\p{1}. This is comparable to the kinetic power of the outflow of molecular gas, which is in the range of $(2 - 18) \times 10^{42}$ erg s\p{1}. 
Therefore, except in the unlikely scenario of a very high efficiency of energy transfer, it is not possible for the radiation to drive the massive molecular gas outflow. The radio jet, on the other hand, can easily act as the driver, since the jet power ($1.3 \times 10^{44}$ erg s\p{1}) is at least an order of magnitude greater than the kinetic power of the outflow \citep{Murthy22b}. Therefore, \target\ provides a unique opportunity to study the impact of radio jets on the host galaxy in detail at a high spatial resolution. 

Here, we present our follow-up CO(2-1) observations using NOrthern Extended Millimeter Array (NOEMA). We expect the strong jet-ISM interaction revealed by the earlier CO(1-0) observations to disturb the gas significantly and thereby alter the physical conditions, compared to the quiescent gas. With the new observations, we have been able to study the gas kinematics in greater detail as well as the physical conditions of the gas via the CO(2-1)/CO(1-0) line ratio.
We used a redshift of $z=0.0165$\footnote{At this redshift, 1$''$ corresponds to 0.33 kpc assuming a flat Universe with H$_{0} = 70$ \kmps\ Mpc$^{-1}$, $\Omega_\Lambda = 0.7$, $\Omega_{\rm M} = 0.3$.}. We describe the observations and data reduction in Sect. 2, report our results in Sect. \ref{sec:r&d}, and discuss the implications and present our conclusions in Sects. 4 and 5.

\section{Observations and data reduction}
\label{sec:observations}
We carried out CO(2-1) observations of \target\ with NOEMA in the D configuration over two observing runs in July and August, 2022 for a total of 7.1 hours. We used 3C\,84 for the bandpass calibration, MW\,349 and B2010+723 for the flux calibration, and B0234+285 and B0307+380 for the phase and amplitude calibration. We used two sidebands, each with a 7.8 GHz bandwidth with 4063 channels, giving a spectral resolution of 1.9 MHz or 2.6 \kmps. The upper sideband was centred at 228.8 GHz, covering the redshifted CO(2-1) frequency of 226.798 GHz, while the lower side band was centred at 213.2 GHz. 

The data reduction steps we followed are described in detail in \citet{Murthy22b}. We used $z=0.0165$ to de-redshift the \textit{uv} data before making the final cube, using natural weighting. The final cube has an RMS noise of 1 mJy beam\p{1}, at  a velocity resolution of 26.3 \kmps, and a restoring beam of $1.59'' \times 1.42''$ with PA$=86.72^\circ$. We extracted the moment maps from this cube shown in Fig.~\ref{fig:moms} using Source Finding Application \citep[SoFiA;][]{Serra15, Westmeier21}\footnote{https://github.com/SoFiA-Admin/SoFiA.}.

\begin{figure*}
\centering
    \includegraphics[width=16cm]{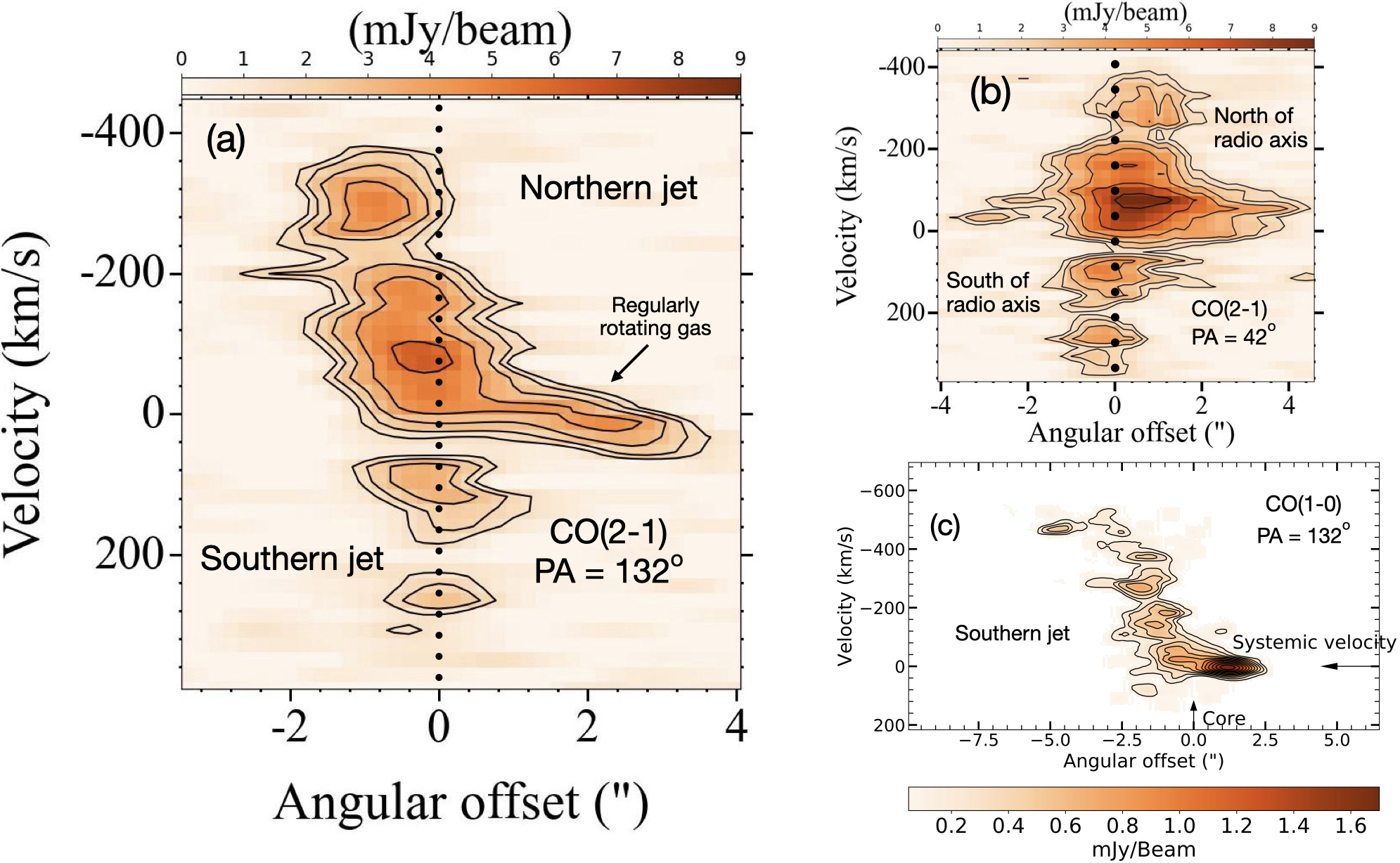}
    \caption{Position-velocity slices of the molecular gas disc. \textbf{(a)}  PV slice along the radio axis. The dashed line shows the location of the core and divides the gas into two regions along the two jets.  We find that the gas with anomalous velocities is offset from the core by \apx500pc along the southern radio jet. \textbf{(b)}  PV slice along the axis perpendicular to the radio jets (PA=$42^\circ$). The dotted line corresponds to the radio axis and divides the gas into regions to the north and south of the radio axis. The blue-shifted gas is concentrated to the north of the radio axis, while the redshifted gas corresponds to the south of the radio axis.
    \textbf{(c)}  PV slice of the gas seen in CO(1-0) emission along the radio axis, where only the blue-shifted gas was detected \citep{Murthy22b}.}
    \label{fig:pvd}
\end{figure*}

The line ratio map presented (Fig. \ref{fig:line_ratio}) is the ratio of the moment-0 maps of CO(2-1) and CO(1-0) transitions. For this purpose, the CO(2-1) cube was convolved with the same restoring beam ($1.9'' \times 1.5''$ with PA=$29.2^\circ$) as that of the CO(1-0). It also has the same velocity resolution of 42.4 \kmps.  To construct a line ratio map, we first obtained a mask for CO(1-0) cube to include emission above the 4$\sigma$ level and used that to obtain a moment-0 map. As we describe in the subsequent sections, we detected more gas in CO(2-1) emission compared to that in CO(1-0). Therefore, to include only the channels and the regions where we detect emission from both transitions, we used the same mask as that used for CO(1-0) emission to obtain a moment-0 map for the CO(2-1) transition. Finally, we took the ratio of the two maps thus obtained. We note that in the regions where only CO(2-1) is detected, the 3$\sigma$ lower limit on the line ratio ranges from 1 to 3.

The CO(1-0) observations had revealed the presence of a regularly rotating CO ring, \apx 10 kpc in size (\apx 30$''$). At the observed CO(2-1) frequency of \apx 229 GHz, it is outside the antenna half-power beam size of \apx 20$''$ and would require wide-field imaging for any further analysis. Hence, we do not discuss it here.

\section{Results}\label{sec:r&d}

\subsection{Distribution and kinematics of the gas}
\label{sec:bubble}

Consistent with our earlier CO(1-0) study, we detect CO(2-1) emission in the central region of NGC~1167. 
The CO(2-1) emission is distributed over a region of \apx7$''$ corresponding to 2.3 kpc. The emission is brightest not at the core but instead in the region overlapping with the southern radio jet, coinciding with where the jet bends strongly (see Fig. \ref{fig:moms}, left)
The velocity field (Fig. \ref{fig:moms}, centre) shows 
the possible presence of two structures. 
A large-scale gradient is observed from NW to SE  (along a position angle close to the radio axis, PA=$132^\circ$) indicating that the gas at large radii follows a relatively regular rotation.
It is worth noting that the position angle of the gas possibly associated with regular rotation is different from the large-scale rotation of the galaxy which has PA = 64$^\circ$ \citep[see Fig. 4, right panel of][]{Struve10c}.
In addition to a rotating structure, the isovelocity contours show a twist in the centre, with the gradient  instead being EW. The difference with respect to the rotating structure suggests the presence of non-radial motion, such as gas outflow. 

The velocity dispersion (Fig. \ref{fig:moms} right) further confirms this. In the outer regions, the dispersion is low, but a region of high velocity dispersion is located just south
of the core, with well above $\sigma = 60$ \kmps\ and up to $\sigma = 180$ \kmps, much higher than that expected for cold molecular gas. A ridge of higher dispersion extending along the north-east is also seen, roughly perpendicular to the direction of the radio axis. This region extends up to about  2$''$ from the radio axis, corresponding to about 0.7 kpc. A similar trend is seen in the CO(2-1)/CO(1-0) line ratio, with higher ratios extending to the north-east of the radio axis (see below). 

The most interesting aspect of the kinematics is revealed by the position-velocity (PV) diagrams. Figure \ref{fig:pvd} shows the presence of gas with large velocities, ranging between about $-360$ \kmps\ and $+250$ \kmps\,, with respect to the systemic velocity of the galaxy. Gas at negative velocities was already found in the previous study of CO(1-0), as shown in panel (c) from \cite{Murthy22b}. The CO(2-1) observations now further reveal the presence of gas with also large redshifted velocities. This is especially clear from the PV slice along the radio axis (PA=132$^\circ$; Fig. \ref{fig:pvd}a). 
This slice gives a view of the actual distribution of the gas and confirms the presence of at least two components as derived from the moment maps. In addition to the  highly blue- and redshifted gas up to \apx 400 \kmps, gas close to the systemic velocity ($<100$ \kmps) is also observed, extending up to almost 4$''$ from the centre. This gas is possibly arising from regularly rotating gas, tracing the gradient in velocity observed in the velocity field of Fig. \ref{fig:moms} (centre). 
The gas with large excursions from the systemic velocity of the galaxy is more prominent in the region of the southern jet and offset from the radio core by \apx 1.5$''$, similar to that seen earlier in CO(1-0) emission (see Fig. \ref{fig:pvd}c).

Thus, the large offset of the  blue- and redshifted components of gas towards the southern radio jet strongly suggests that they do not belong to a regularly rotating structure. The gas around the systemic velocity appears more symmetric about the core, with a lower velocity dispersion; hence, it is likely to be rotationally supported. 

The PV diagrams further support this scenario. Figure \ref{fig:pvd}b shows the PV in the direction perpendicular to the radio jet (along 42$^\circ$). If the circumnuclear gas does indeed have a component of rotation with its major axis close to the radio axis, this perpendicular direction would roughly represent the minor axis. Thus, for regularly rotating gas, we would expect the velocities of the gas to be close to the systemic velocity. Indeed, some of the gas follows this prediction, confirming the presence of a rotating structure. However, we also find that a large fraction of the gas deviates from this. In particular, we detected gas with large velocities deviating from the systemic velocity.  Interestingly, we note that the location of the red- and blue-shifted gas is offset by \apx 1-1.5$''$ in opposite directions: the blue-shifted gas is located to the north of the radio axis, while the redshifted gas to the south. This suggests they are tracing gas in different regions. This aspect will be important for the interpretation of the geometry of the system (see Sect. \ref{sec:d&c}).

\begin{figure}
\centering
    \includegraphics[width=8.5cm]{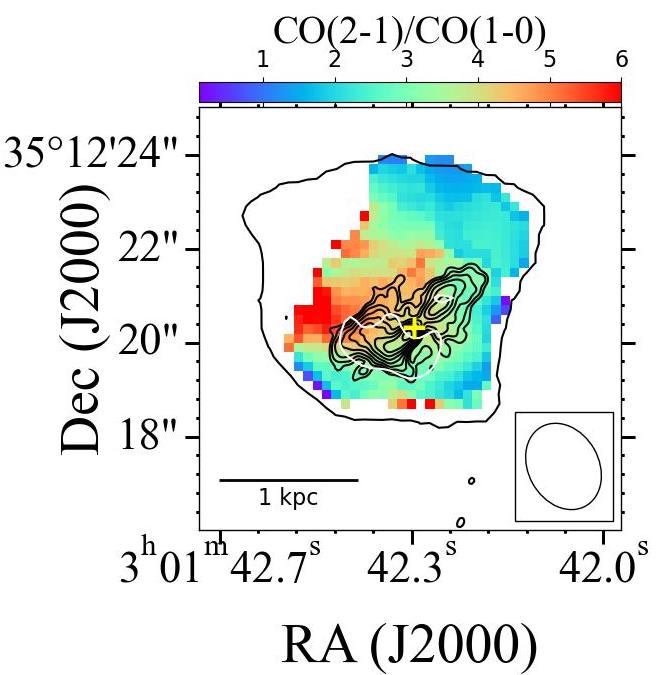}    
    \caption{The CO(2-1)/CO(1-0), line ratio (${\rm R_{21}}$) map for the regions where the emission is detected in both transitions. The black contour marks the entire region from which we detect CO(2-1) emission. The black contours in the centre show the 8.4 GHz radio continuum from \citet{Giroletti05}. The yellow cross shows the radio core. The white contour shows the region of soft X-ray emission detected in  \textit{Chandra} observations reported in \citep[see Fig. 6, right, in][]{Fabbiano22}.}
    \label{fig:line_ratio}
\end{figure}

\begin{figure*}
\centering
    \includegraphics[height=7cm]{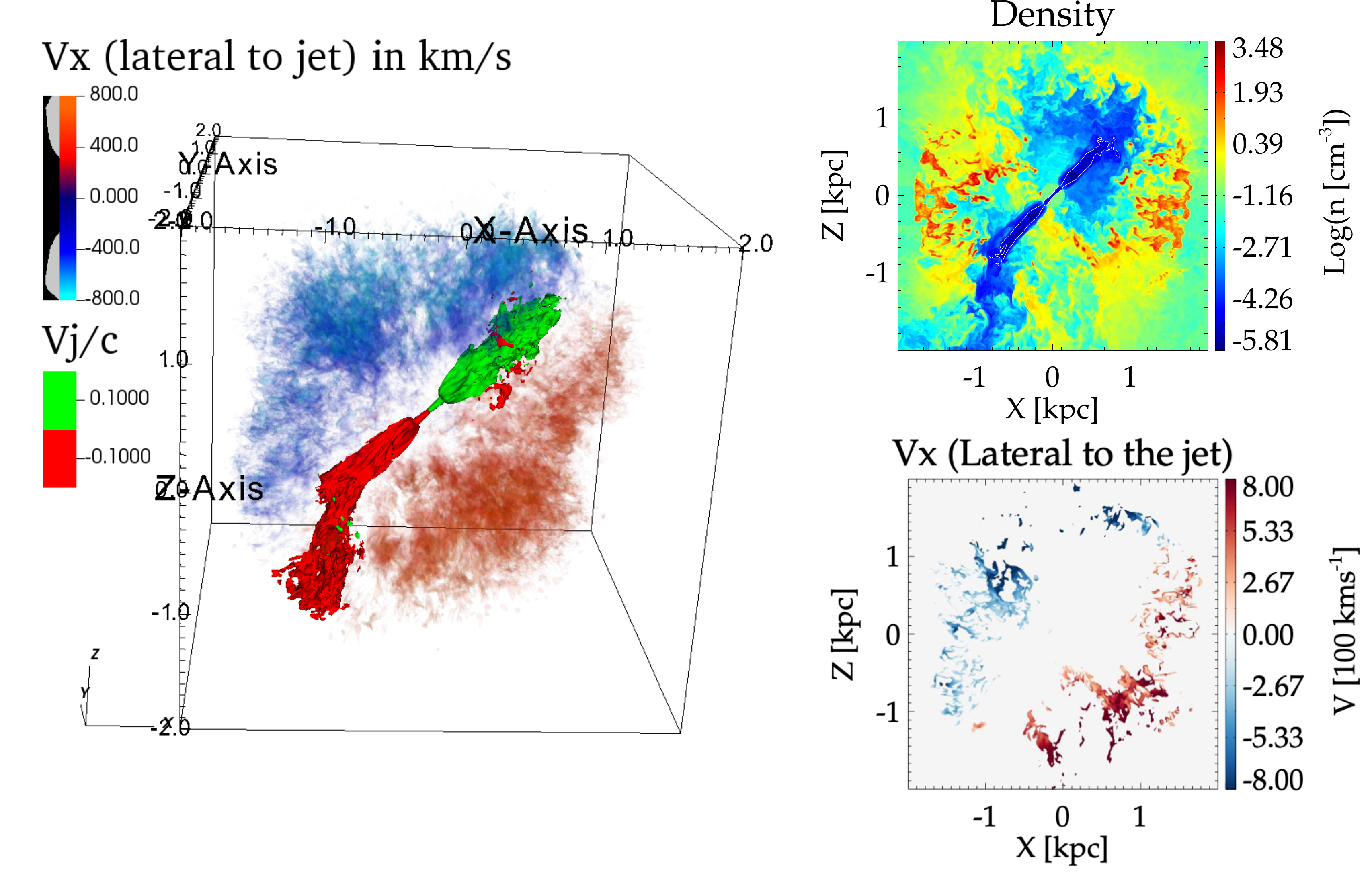}
    \includegraphics[height=6cm]{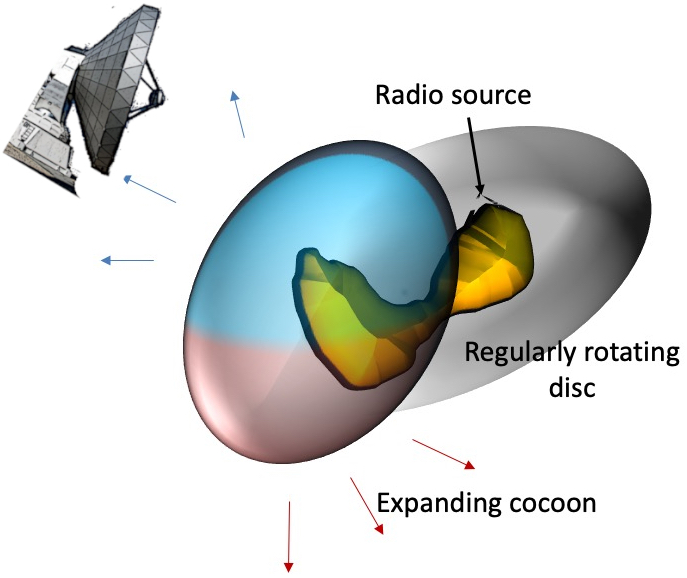}
    \caption{\textbf{Left}: Panel (a) depicts the 3D volume rendering of the velocity component lateral to the jet axis (see Sec.~\ref{sec:d&c} for details) for dense gas ($n \geq 10 \mbox{cm}^{-3})$, from the simulations of jet-ISM interaction in \citet{Fabbiano22}. A 2D cross-section of this image in the $X-Z$ plane is shown in panel (c). The red and blue colours show oppositely directed flows with respect to the jet-axis, expanding as a bubble. The pair of relativistic jets are presented using the contours of $|v| = 0.1$ c in green and magenta. Panel (b) shows the cross-section of the gas density in $X-Z$ plane. \textbf{Right}: Cartoon image representing the proposed model of the evolution of a jet-driven bubble in the southern jet, where the line of sight traces both sides of the expanding bubble.
    }
    \label{fig:cartoon}
\end{figure*}

\subsection{Physical conditions of the gas}
\label{sec:line-ratio}

With the two CO transitions, we  derived an image of the CO(2-1)/CO(1-0) flux ratio, ${\rm R_{21} \equiv S_{\rm CO(2-1)}/S_{\rm CO(1-0)}}$ (Fig. \ref{fig:line_ratio}). The image clearly shows the presence of structure in the distribution of the line ratio. Here, ${\rm R_{21}}$ ranges from 1.4 to beyond 5. These values indicate the presence of both low-excitation ($<<4$) and high-excitation components ($>4$).  The largest 
ratios are observed in the region south of the core, starting from where the southern jet bends sharply, following the structure of the bent jet. In this region, ${\rm R_{21}} \sim 5$, well above the expected values for quiescent molecular gas \citep[e.g.][]{Bolatto13,Oosterloo17}. Although it is not as extreme as found in other radio galaxies \citep[e.g.][]{Oosterloo17, Oosterloo19, Morganti21}, it is still beyond the limit of 4 (or 1 in brightness temperature ratios) for optically thick emission. This implies that the gas is likely thermalised: it is highly dense, optically thin, and characterised by a high excitation temperature. The spatial coincidence of this region of high R$_{21}$ with the bent part of the radio jet indicates that the high excitation of the gas is likely due to the dissipation of the jet kinetic energy.

The high ${\rm R_{21}}$ also explains the non-detection of the redshifted high velocity component in CO(1-0): the faintness of this transition resulted in the emission being below the detection limit of the observations. Despite the low S/N of the CO(1-0) emission, we can still infer that the region of high excitation covers at least a part of the region with high velocity dispersion. Also, to first order, it is also elongated along the same direction (see Fig. \ref{fig:moms}, right), perpendicular to the southern jet up to \apx 2$''$, corresponding to 0.7 kpc from the radio axis. This also corresponds to the region where the southern radio jet bends sharply, suggesting that the kinematically disturbed gas is also of high excitation. This is in agreement with the findings in other galaxies, where excitation temperatures in the range 20–100 K are found \citep[e.g.][]{Oosterloo17,Oosterloo19,Ruffa20,Morganti21,Audibert23}.

\subsection{Outflow mass and energetics}
\label{sec:energetics}

Given the high R$_{21}$ in addition to the complex gas kinematics, we needed to revise the outflow gas mass and the outflow rate estimated in \citet{Murthy22b}. Based on our new findings, we conclude that the CO to H$_2$ conversion factor, $\alpha = 0.34$ \Msun (K \kmps\ pc\pp{2})\p{1} corresponding to highly turbulent, optically thin gas, is the most suitable assumption for the kinematically disturbed gas. In CO(2-1) emission, we detected more gas in the redshifted part of the expanding bubble. This contributes to $\sim 25\%$ of the flux of the outflow. Thus, we infer a revised outflow gas mass, M$_{\rm H_2} \sim 6.3 \times 10^6$ \Msun\ for $\alpha = 0.34$ \Msun (K \kmps\ pc\pp{2})\p{1}. 

We derived the mass outflow rate $\dot{M}_{\rm out} = M_{\rm H_2}/\tau_{\rm dyn}$, where $\tau_{\rm dyn}$ is the time taken by the outflowing gas to reach the current location $r_{\rm out}$. $\tau_{\rm dyn}$ in turn depends on the outflow velocity. 
Various approaches have been used to derive these parameters, focussing, in particular, on the selection of the outflow velocity (see e.g. \citealt{Harrison18} for an overview). We estimated the $\dot{M}_{\rm out}$ for two extreme cases: (a) $v_{\rm out} = 360$ \kmps, which is the velocity of the far wings of the disturbed gas (Fig. \ref{fig:pvd}), and, therefore, the maximum velocity at which the gas is observed; (b) a more conservative estimation of $v_{\rm out} = 200$ \kmps, the velocity at which the bulk of the outflow is observed, based on a visual inspection of the PV plots (Fig. \ref{fig:pvd}). In the latter case, the higher velocities of the gas are assumed to be due to the turbulence in the flow. Based on our findings in Sect. \ref{sec:bubble}, we assume $r_{\rm out}=0.7$ kpc and obtain a mass outflow rate in the range \apx 1.8 to \apx 3.2 \Msun yr\p{1} for $\alpha = 0.34$ \Msun (K \kmps\ pc\pp{2})\p{1}. 

It is interesting to derive the kinetic energy associated with the outflow and compare it with the jet power. The latter was estimated in the range log P$_{\rm  jet} = 43.91$ and 44.13 erg s$^{-1}$, using the relations obtained by \citet{Willott99} and \citet{Cavagnolo10}, respectively, as described in \citet{Murthy22b}. To derive the kinetic power of the outflow, we used the relation:  ${\Dot{\rm E}_{\rm kin}} = (\Dot{\rm M}_{\rm out}/2) \times (v_{\rm out}^2+3*\sigma^2)$.  As $\sigma$, we use an average value of 150 \kmps\ from the outflowing region in Fig. \ref{fig:moms} right panel. The derived values of ${\Dot{\rm E}_{\rm kin}}$ for the two values of $v_{\rm out}$ described above range between $6.1 \times 10^{40}$ and $2.0 \times 10^{41}$ erg s$^{-1}$. 
Thus, we confirm the conclusion of \citet{Murthy22b} that despite the low power, the radio jet is capable of driving the observed molecular gas outflow, even for a low coupling efficiency.

\section{An expanding cold gas bubble driven by the jet}
\label{sec:d&c}

The new CO(2-1) observations provide a more complete picture and a better characterisation of the ongoing jet-ISM interaction in \target, expanding the results from the CO(1-0) observations alone (\citet{Murthy22b}). The presence of rotating gas as well as gas with velocities up to almost 400 \kmps\ deviating from regular rotation is now much clearer. The blue- and redshifted gas is detected mostly along the southern radio jet. Furthermore, we found that the blue-shifted component is concentrated to the north of the radio axis, while the redshifted component is concentrated to the south of the radio axis (see Fig. \ref{fig:cartoon}). 

These characteristics are consistent with the red- and blue-shifted gas components tracing the interaction between the jet and the ISM and the expansion resulting from this. The presence of an such interaction is further confirmed by the finding of high ${\rm R_{21}}$ values ($>4$) in the region around the southern jet and the location of the jet bending (perpendicular). 

In our earlier \textit{Chandra} X-ray study of this source, we found that the soft X-ray emission envelops the southern radio jet and extends perpendicularly to the radio axis \citep[see Fig. 6, right, in][]{Fabbiano22}, similar to the trends seen in velocity dispersion and R$_{21}$, indicating the presence of dense gas and strong shocks. \citet{Fabbiano22} suggested that the soft X-ray emitting gas covers the molecular gas from the outside. Therefore, we conclude that the expanding bubble consists of multi-phase gas. 

Our earlier studies of \target\ successfully confirmed several predictions of the simulations regarding the impact of low-power radio jets on the ambient ISM \citep{Murthy19, Murthy22b, Fabbiano22}. In these studies, we presented qualitative comparisons of our observed results with the simulations of a jet \citep{Mukherjee18b} expanding into a clumpy gas disc at an angle of 45$^\circ$. We had noted that the observed velocities seen in our CO(1-0) observations \citep[][]{Murthy22b} were reproduced in the simulations, as was the observed morphology of the radio source. Additionally, the simulations also showed the presence of redshifted gas, as well as rotating circumnuclear disc, which were not clearly detected in those observations. Such comparisons were further extended in \citet{Fabbiano22}, where new simulations of a jet, of power $10^{44}$ erg s\p{1}, were presented to explain the observed emission of X-rays from shocked gas. 

We revisited the simulations presented in \citet{Fabbiano22} to demonstrate that a low power jet interacting with a gas disc can blow out a large-scale bubble, whose oppositely expanding walls are likely being observed in this current study. The left panel of Fig. \ref{fig:cartoon} shows a 3D volume rendering of the velocity component perpendicular to the jet-launch axis\footnote{The velocity component $v_x$ presented in Fig.~\ref{fig:cartoon} is obtained by rotating the coordinate system to orient the $Z$ axis to be along the jet-launch axis, which is $45^\circ$ with respect to the disk normal.} for the dense gas ($n \geq 10~\mbox{cm}^{-3}$ and dense gas tracer value $\geq 0.1$) in blue and red. A 2D slice of the density and the said lateral velocity in the $X-Z$ plane are presented in the middle panel.  We find that the jet--ISM interaction indeed results in the creation of an expanding bubble, visible as blue- and redshifted fronts on either side of the radio jets, matching the observations. Such an expanding gas structure is created as a result of the partial isotropisation of jet energy in the strong jet--cloud interactions. This is expected to happen through 1) the creation of multiple split jet streams, in addition to the main bent jet stream, that flow in all directions and 2) the thermalisation of the jet kinetic energy through strong shocks around the clouds and the resulting isotropic momentum boost. Hence, the simulations support our proposed scenario, presented as a cartoon in the right panel of Fig. \ref{fig:cartoon}, namely, that our current observations are tracing the two halves of an expanding bubble at the southern jet. 

\section{Implications for galactic-scale feedback from radio AGN}

With our study of \target, a low-accretion, low-luminosity optical AGN, we have been able to isolate the impact of a low-power jet on the surrounding medium. The coupling efficiency, defined as the ratio of the kinetic energy associated with the outflow to the jet power, is 0.1\% -- 1\% in this case (see Sect. \ref{sec:energetics}). This implies that most of the energy in the jet is not used to drive the outflow and will likely be deposited at larger distances.

Notably, a signature of this phenomenon has been detected in \target. An integral field unit (IFU) study by \citet{Gomes16a} tracing ionised gas in the central region of this galaxy has found that the line ratios in the central region (with a 1.4 kpc diameter, perpendicular to the radio axis) fall in the LINER part of the BPT diagram; furthermore, the high equivalent width of the \halpha\ lines in this region indicate that the evolved pAGB stars alone cannot be the source of ionisation \citep[see Fig. 1 of][]{Gomes16a}. Combining this with our findings about cold gas, and X-ray studies, we further conclude that this high-\halpha    \  equivalent-width gas is also part of the expanding multi-phase gas bubble and. As such, this is evidence of the far-reaching effect of jet--ISM interaction. 

Therefore, this is one of the few cases where the impact of the jet is observationally confirmed to be far-reaching and not just along the path of the radio jets. Such an impact has also been observed in other sources with low-power jets \citep[e.g.][]{Audibert23,Girdhar24}) and also high-$z$ galaxies \citep[e.g.][]{Roy24, D'Eugenio24}. However, these sources are also high in radiation, making it difficult to disentangle the effect of radio jets from the former, unlike in the case of \target. 

We further confirm some of the key predictions of the numerical simulations \citep[e.g.][]{Sutherland07, Mukherjee16, Mukherjee18b}: low-power radio jets, when expanding into circumnuclear gas discs impact the ISM strongly; such a jet--ISM interaction results in kinematically disturbed gas even at distances beyond the apparent extent of the radio jets. Furthermore, we also confirm the kinematic signatures of such jet--ISM interactions seen in synthetic velocity fields and PV diagrams of these simulations. While the observational evidence is building up, the remarkable similarities already serve to strongly support the prediction of these models that kpc- and sub-kpc jets do impact the gas in their host galaxies over spatial scales larger than their apparent sizes. 

Low-power radio sources (L$_{\rm1.4 GHz} \leq 10^{23}$ W Hz\p{1}) form the majority of the radio AGN population \eg{Best05, Sabater19}. Furthermore, it has recently been found that there is an excess of kpc- and sub-kpc-scale radio sources in flux-limited samples \citep[e.g.][]{Odea21, Readhead24}, suggesting that a large fraction of the radio sources will remain embedded within their host galaxies for the majority of their lifetime. In this context, our findings that such kpc-scale, low-power sources are capable of having a significant impact on the gas within their host galaxies highlight the importance of these sources in galaxy evolution, and as such, suggest their inclusion in cosmological models.


\begin{acknowledgements}

We thank the referee for constructive comments on the paper. SM thanks Michael Bremer for help with data reduction. SB thanks the EU Erasmus$+$ program of the European Union for support during the project. We thank Marcello Giroletti for providing us with the 8.4 GHz VLA radio continuum image. This work is based on the observations carried out under project number S22BJ with the IRAM NOEMA Interferometer. IRAM is supported by INSU/CNRS (France), MPG (Germany) and IGN (Spain). We thank the IRAM staff for making these observations possible.

\end{acknowledgements}

%
   \bibliographystyle{aa} 
   \bibliography{ref} 
%


\end{document}